\documentclass{article}%
\usepackage{amsmath}
\usepackage{amsfonts}
\usepackage{amssymb}
\usepackage{graphicx}%
\providecommand{\U}[1]{\protect\rule{.1in}{.1in}}

\begin{document}

\title{Localization and the interface between quantum mechanics, quantum field theory
and quantum gravity II\\(The search of the interface between QFT and QG)\\{\small dedicated to the memory of Rob Clifton}\\{\small to be published in "Studies in History and Philosophy of Physics"}}
\author{Bert Schroer\\CBPF, Rua Dr. Xavier Sigaud 150 \\22290-180 Rio de Janeiro, Brazil\\and Institut fuer Theoretische Physik der FU Berlin, Germany}
\maketitle
\tableofcontents

\begin{abstract}
The main topics of this second part of a two-part essay are some consequences
of the phenomenon of \textit{vacuum polarization} as the most important
physical manifestation of \textit{modular localization.} Besides
philosophically unexpected consequences, it has led to a new constructive
"outside-inwards approach" in which the pointlike fields and the compactly
localized operator algebras which they generate only appear from intersecting
much simpler algebras localized in noncompact wedge regions whose generators
have extremely mild almost free field behavior.

Another consequence of vacuum polarization presented in this essay is the
localization entropy near a causal horizon which follows a logarithmically
modified area law in which a dimensionless area (the area divided by the
square of dR where dR is the thickness of a light sheet) appears. There are
arguments that this logarithmically modified area law corresponds to the
volume law of the standard heat bath thermal behavior. We also explain the
symmetry enhancing effect of holographic projections onto the causal horizon
of a region and show that the resulting infinite dimensional symmetry groups
contain the Bondi-Metzner-Sachs group. When the first version of this paper
was submitted to hep-th it was immediately removed by the moderator and placed
on phys. gen without any possibility to cross list, even though its content is
foundational QFT. With the intervention of a member of the advisory comitee at
least the cross-listing hopefully seems now to be possible.

\end{abstract}

\section{Introduction to the second part}

Whereas the first part \cite{interI} presented the interface between
(relativistic) QM and QFT, this second part focusses on the interface of QFT
with gravity, or more precisely on what hitherto was presumed to define this
interface\footnote{An example is the concept of entropy associated with
horizons, which according to our results in this essay can be formulated and
understood in the standard setting of LQP.}. As a result of the very different
nature of the localization concept of QFT, a bipartite partition into a
subalgebra of a causally closed region and its causal disjoint does not tensor
factorize; the sharp localization generates infinitely large vacuum
polarization which destroys the quantum mechanical entanglement concept. This
is however not the end of the story, since in QFTs with reasonable phase space
degrees of freedom one can enforce a tensor factorization via the split
construction which re-creates some but not all aspects of quantum mechanical entanglement.

The next section presents this important "splitting" idea. Although there are
mathematical examples of QFT which violate the prerequisites for splitting, a
physically motivated phase space density of QFT exclude such cases. In
particular the existence of the completeness property of asymptotic incoming
or outgoing particles in theories with a finite number of particle species
imply the splitting property. The third section addresses the most important
physical implication of localization, namely \textit{vacuum polarization} and
prepares the ground for the presentation of localization entropy in section 4.
It is shown that if in an interacting theory one "bangs" with a local operator
$A$ onto the vacuum, the so-obtained local vacuum excitation state $A\Omega$
has infinitely many particle/antiparticle components whose analytic
continuation determine all formfactors of $A$ through the crossing relations.
Section 5 explains the mathematical/conceptual meaning of holography onto
horizons and shows how the loss in information and spacetime symmetry can be
reconciled with a huge in conformal symmetry: the holographic projection
admits infinite dimensional symmetry groups which contains in particular the
classical Bondi-Metzner-Sachs group \cite{BMS}. The increase of conformal
symmetry on the horizon does not help in inverting the holographic projection
back towards the reconstruction of the bulk.

The perhaps greatest progress has been the adaptation of the Einstein local
covariance principle to the quantum realm \cite{BFV} which is briefly sketched
in section 6. This goes a long way towards the "background independence" which
is the would-be "Holy Grail" of the still elusive QG.

The modular theory of which relevants parts for the 2-part paper were
presented in part I, plays an important role in connecting the entropy aspects
of the heat-bath situation with those of localization. According to our best
knowledge the connecting formulae in section 4 are new.

\section{The split inclusion}

There is one property of LQP which is indispensable for understanding how the
quantum mechanical tensor factorization can be reconciled with modular
localization: the \textit{split property}.

\textbf{Definition: }\textit{Two monads }$\mathcal{A},\mathcal{B}$\textit{ are
in a split position if the inclusion of monads }$\mathcal{A}\subset
\mathcal{B}^{\prime}$\textit{ admits an intermediate type I factor
}$\mathcal{N}$\textit{ such that }$\mathcal{A\subset N\subset B}^{\prime}$

Split inclusions are very different from modular inclusions or inclusions with
conditional expectations. Their main property is the existence an
$\mathcal{N}$-dependent unitarily implemented isomorphism of the
$\mathcal{A},\mathcal{B}$ generated operator algebra into the tensor product
algebra%
\begin{equation}
\mathcal{A}\vee\mathcal{B\rightarrow A}\otimes\mathcal{B\subset N}%
\otimes\mathcal{N}^{\prime}=B(H) \label{split}%
\end{equation}
The prerequisite for this factorization in the LQP context is that the monads
commute, but it is well-known that local commutativity is not sufficient, the
counterexample being two double cones which touch each other at a spacelike
boundary \cite{Haag}. But as soon as one localization region is spacelike
separated from the other by a (arbitrary small) spacelike security distance,
the interaction-free net satisfies the split property under very general
conditions. In \cite{BDF} the relevant physical property was identified in
form of a phase space property (part I, section 4). Unlike QM, the number of
degrees of freedom in a finite phase space volume in QFT is not finite, but
its infinity is in some sense mild; it is a \textit{nuclear set} for free
theories and this nuclearity requirement\footnote{A set of vectors is nuclear
if it is contained in the range of a trace class operator.} is then postulated
for interacting theories \cite{Haag}. The physical reason behind this
nuclearity requirement is that it allows to show the existence of temperature
states once one knows that a QFT exists in the vacuum representation. Even
more important its validity prevents the violation of the causal shadow
property which states that the degrees of freedom in the causal shadow of a
spacetime region are the same as those in the original region: $\mathcal{A(O}%
\text{\textquotedblright}\mathcal{)=A(O}$) which is the algebraic analog of
hyperbolic Cauchy propagation. All these properties are formally true in a
Lagrangian setting but they constitute a physical safety kit for movements
outside the standard quantization parallelism to classical field theory as
holography onto horizons, AdS-CFT correspondences etc.

The split property for two securely causally separated algebras has a nice
physical interpretation. Let $\mathcal{A}=\mathcal{A(O}),~\mathcal{B}^{\prime
}=\mathcal{A(\check{O}}),~\mathcal{O\subset\check{O}}.$ Since $\mathcal{N}$
contains $\mathcal{A}$ and is contained in $\mathcal{B}^{\prime}$ (but without
carrying the assignment of a sharp localization between $\mathcal{O}$ and
$\mathcal{\check{O})}$, one may imagine $\mathcal{N}$ as an algebra which
shares the sharp localization with $\mathcal{A(O})$ in $\mathcal{O}$, but its
localization in the "collar" between $\mathcal{O}$ and $\mathcal{\check{O}}$
is "fuzzy" i.e. the collar subalgebra is like a "haze" which does not really
occupy the collar region. This is precisely the region which is conceded to
the vacuum polarization cloud in order to spread and thus avoid the infinite
compression into the surface of a sharply localized monad. If we take a
sequence of $\mathcal{N}^{\prime}s$ which approach the monad $\mathcal{A},$
the vacuum polarization clouds become infinitely large in such a way that no
direct definition of e.g. their energy or entropy content is possible.

The inclusion of the tensor algebra of monads into a type I tensor product
(\ref{split}) looks at first sight like a d\'{e}j\`{a} vu of QM tensor
factorization, but there are interesting and important differences. In QM the
tensor factorization obtained from the Born localization projector and its
complement is automatic since the vacuum of QM (or the ground state of a
quantum mechanical zero temperature finite density system) tensor factorizes.
In QFT the vacuum does not tensor factorize at all, but there are other states
the so-called "split vacuum" states in the Hilbert space which emulate a
tensor-factorizing vacuum in the sense that expectation values of operators in
$\mathcal{A(O})\vee\mathcal{A(\check{O}}^{\prime})$ factorize in the split
vacuum%
\begin{equation}
\left\langle 0_{split}\left\vert AB\right\vert 0_{split}\right\rangle
=\left\langle 0\left\vert A\right\vert 0\right\rangle \left\langle 0\left\vert
B\right\vert 0\right\rangle ,~A\in\mathcal{A(O}),B\in\mathcal{A(\check{O}%
}^{\prime})
\end{equation}
but there remains a huge conceptual difference to the quantum mechanical Born
factorization of the "nothing" state. The splitting process requires the
supply of energy since the split vacuum has infinite vacuum polarization (with
finite mean energy) in the collar region which is spacelike to $\mathcal{O}%
\vee\mathcal{\check{O}}^{\prime}.$ The physical states of QFT are by
definition the states with arbitrary large but finite energy. Their massive
particle content is finite but they may contain (as it is the case in QED)
infinitely many zero mass particles. In contradistiction to QM these states
only tensor-factorize after a spatial split, in which case the reduced vacuum
and all finite energy states become thermal Gibbs state with respect to a
split-related Hamiltonian. Without the split the $\mathcal{A(O})$-reduced
vacuum state is a singular\footnote{A singular KMS state denotes a KMS state
which is not a Gibbs state.} thermal KMS state.

The problem of physical realizability has not been given much attention in
foundational discussions of QM. In QFT this issue is more serious since the
situations are much more counter-intuitive, as was shown before with the
particle behind the moon argument for the global vacuum. This property is
absent in a split vacuum state; the split defines a barrier, but it is unclear
how such split states can be prepared and monitored.

Most foundational properties of QM, as violation of Bell's inequalities, the
Schroedinger cat property and many other strong deviations from classical
reality can be experimentally verified. This is generally not possible for the
vacuum polarization caused properties which result from modular localization
simply because macroscopic manifestations are too small. A typical example is
the Unruh effect i.e. the thermal manifestation of a uniformly accelerated
particle counter in the global vacuum, where the temperature created by an
acceleration of 1m/sec is $10^{-19}K$ too small for ever being registered. But
for the perception of the reality which underlies LQP the difficulty in
registering such effects does not diminish their conceptual importance.

The characterization of the restriction of the global vacuum to a local
algebra in terms of a thermal state for a modular Hamiltonian holds,
independent of whether the local algebra is a sharply localized monad
$\mathcal{A(O)}$ or a type I factor $\mathcal{N}$ contained in a larger
sharply localized algebra $\mathcal{\check{O}}$, as in the above splitting
construction. The only difference is that the in the second case the KMS state
is also a Gibbs state i.e. the Hamiltonian on $\mathcal{N}$ has a discrete
spectrum (in case $\mathcal{\check{O}}$ is compact). This thermal
reinterpretation of reduced states does not only hold for the vacuum, but
applies to all states which are of physical relevance in particle physics i.e.
to all finite energy states for which the Reeh-Schlieder theorem applies.

Since KMS states on type I factors are Gibbs states, there exists a density
matrix. Therefore these Gibbs state can have a finite energy and entropy
content which for monads is impossible. But a monad may be approximated by a
sequence of type I factors in complete analogy to the thermodynamic limit. In
fact the thermodynamic limit is the only place where a monad algebra appears
in a QM setting; an indication that this limit is accompanied by a qualitative
change is the fact that one looses the density matrix nature of the Gibbs
state which changes to a more singular KMS state which simply does not exist
on quantum mechanical type I algebras. A related fact is the breakdown of the
tensor factorization into physical degrees of freedom and their "shadow world"
in the thermodynamic limit. This implies that the "Thermofield formalism" is
only applicable for the boxed QFT (type I), it looses its meaning in the
thermodynamic limit when the state becomes a singular KMS state and the
algebra turns into a monad.

The structural difference can be traced back to the nature of modular
Hamiltonians. Whereas for a monad the modular Hamiltonian has continuous
spectrum\footnote{A typical example is the thermodynamic V$\rightarrow\infty$
limit in which the discrete spectrum of the Hamiltonian in a box turns into a
continuous spectrum and the Hamiltonian becomes the modular Hamiltonian on a
monad. Gibbs state.} and hence an ill-defined (infinite) value of energy and
entropy, this is not the case for the $\mathcal{N}$-associated density matrix
constructed from the split situation. So the way out is obvious, one must
imitate the thermodynamic limit by constructing a sequence of type I factors
(a "funnel" $N_{i}\supset\mathcal{A(O})$ by tightening the split) which
converge from the outside towards the monad; equivalently one may approximate
from the inside.

In the next section the split limit, in which a double-cone localized monad is
approximated by a sequence of type I$_{\infty}$ factors $\mathcal{N}_{i},$
will be presented. In this case the modular group of ($\mathcal{N}_{i}%
,\Omega)$ leads to a Gibbs situation i.e. the restriction of the vacuum to the
algebra $\mathcal{N}_{i}$ is a Gibbs state at the same modular temperature as
the that associated with the restriction to the monad. The main distinction to
the standard heat bath situation is that Hamiltonians which result from
restricting the vacuum to an $\mathcal{N}_{i}$ obtained from the split
construction is constructed according to very different principles from that
of Gibbs states in a finite quantization box with the help of the Hamiltonian
which describes the time development in a quantum mechanical inertial frame.
Nevertheless for two-dimensional chiral theories there exists a rigorous
relation between the two kinds of thermal behavior: \textit{the inverse Unruh
effect}. The localization-caused thermal manifestation in a chiral theory is
related by a conformal trnaformation to the thermodynamic limit of a
one-dimensional global heat bath inertial system \cite{S1}\cite{S2}. There are
presently no reliable computational techniques for dealing with modular
Hamiltonians and their split approximands, although there is no lack of
mathematical precision in \textit{defining} these objects. In the next section
we will try to overcome this situation by invoking geometrical arguments
leading to a localization of the vacuum polarization cloud inside a
\textit{light-sheet}.

The split property does not hold in all axiomatic models of LQP but there are
rather good arguments that it is valid in "physical" models which share a
physically relevant property with Lagrangian models. What is important is not
the Lagrangian quantization but rather the causal timelike propagation aspect
which historically has been referred to as the time-slice property \cite{H-S}
or the causal shadow property. This property is violated in theories with too
many phase space degrees of freedom.

The the study of what constitutes a physical phase space density started in
the 60s \cite{Ha-Sw} when for the first time it was shown that the
implementation of the relativistic causality principle requires a bigger
cardinality of phasespace degree of freedoms than in QM. Whereas the number of
degrees of freedom per unit cell in phase space is finite, these authors found
the cardinality in QFT is infinite in the sense of being a compact set. This
result was later sharpened to $nuclear<compact$ \cite{Bu-Wi} and various
important properties were shown to be consequences, among them the existence
of thermal states for all temperatures and the splitting property
\cite{Do-Lo}. The implies the absence of an unwanted Hagedorn temperature as
it occurs in infinite component QFT with too many degrees of freedom as string
theories. It is easy to invent QFTs which violate this bounds in the sense of
having too many degrees of freedom, the generalized free fields with suitably
increasing Kall\'{e}n-Lehmann density are the simplest examples.

Unfortunately the knowledge about these concepts got lost, otherwise how can
one explain that a worldwide community works on a narrow subject as the
AdS$_{5}$-CFT$_{4}$ correspondence with more than 6000 contributions without
becoming aware that it is structurally impossible to have theories with
physical degrees of freedom on both sides of this correspondence? The naive
expectation (backed up by unconvincing calculations outside any mathematical
control) in these papers are in contradiction with established
facts\footnote{The sharing of the SO(4,2) conformal symmetry prevents a
dilution of degrees of freedom in passing from the higher- to the lower-
dimensional theory. Starting from a standard conformal QFT (the supersymmetric
N=4 Yang-Mills theory is a potential candidate) the degrees of freedom
transplanted to the higher dimensional AdS spacetime are not sufficient to
"fill" the AdS bulk and remain hovering at the boundary.}. One can safely
assume that this knowledge has been lost and even worse, that those few
correct papers which have been published on this correspondence \cite{Re1}%
\cite{Re2}\cite{Du-Re} are beyond conceptual reach of the typical particle
physicist with a contemporary (post string theory) quantum field theoretical
background. If one looks at the thousands of publication on this matter one
cannot escape the conclusion that the deep conceptual knowledge of the past
has been replaced by a more metaphoric mode of thinking in the pursuit of an
ultra-reductionist theory of everything (TOE).

In the above form the positioning of monads aims at characterizing LQP in
Minkowski spacetime. This begs the question whether there is a generalization
to curved spacetime (CST). A very special exploratory attempt in this
direction would be to investigate the Diff(S$^{1}$) symmetries beyond the
Moebius group in chiral theories for their possible modular origin in terms of
positioning monads relative to reference states which are different from the
vacuum. Since the \textit{extended} chiral theories which result from
null-surface holography (and not from chiral projections of a two-dimensional
conformal QFT) seem to have great constructive potential, this question could
be of practical interest.

I expect that by pursuing the algebraization of QFT in CST according to the
positioning of monads viewpoint, one will learn important lessons about the
still unknown QFT/QG interface. It would be of great interest to understand
whether the isometric isomorphisms related to the local covariance principle
(see last section) have modular roots similar to symmetries of the vacuum in
Minkowski QFT. A conservative approach which explores unknown aspects of QFT
while staying firmly rooted in known principles seems to be the most promising
path for pushing the borderline of QFT in CST further towards the still
unknown QFT-QG interface.

\section{Localization-induced vacuum polarization, present view and history}

The phenomenon of vacuum polarization has been the point of departure of many
metaphors of which the vacuum in QFT as a "steaming broil" is perhaps the best
known because it occasionally even entered textbooks. In order to support this
image the appeal to a short time violation of the energy conservation allowed
by the uncertainty relation was made. A less metaphoric view comes from
locally "banging" on the vacuum i.e. applying a compactly localized operator
to the vacuum state. \ Such a state is characterized by its n-particle matrix
elements for all n and these n-particle vacuum polarization components are in
turn special boundary values of an analytic n-particle master function whose
different out-in particle distributions obtained from the vacuum polarization
component by crossing are the formfactors of $A.$
\begin{align}
&  A\Omega\simeq\{\left\langle 0\right\vert A\left\vert p_{1},...p_{n}%
\right\rangle ^{in}\}_{n}\label{form}\\
&  \overset{cros\sin g}{\rightarrow}\{^{out}\left\langle -\bar{p}%
_{k+1},...-\bar{p}_{n}\right\vert A\left\vert p_{1},...p_{k}\right\rangle
^{in}\}_{n}\nonumber
\end{align}
where the negative mass shell momenta $-\bar{p}$ denotes the analytic
continuation which is part of the crossing process and the bar is the reminder
that the particle is an antiparticle to the original particle (this can be
omitted in case of self-conjugate particles). In case of $A\in\mathcal{A(O})$
being the identity operator there is no "banging" onto the vacuum; in that
case we are dealing with the S-matrix for which many matrix-elements vanish as
a result of the energy momentum conservation between the particles without the
feeding in from the localized operator $A.$ In particular the S-matrix is free
of vacuum polarization clouds, since the identity leaves the vacuum invariant
and does not bang. \ 

The crossing property is one of the deepest characteristics linking fields and
particles. It had been known for a long time that there are no compactly
localized operators which applied to the vacuum generate a one-particle state
without an admixture of a vacuum polarization cloud except if the theory is
that of a free field\footnote{Within the Wightman setting this has been known
as the Jost-Schroer theorem \cite{St-Wigh}. A stronger form was recently
proven in the algebraic setting \cite{Mund}.}. One-particle operators exist as
in- or out- operators (or unitary transforms thereof) in the full algebra
$B(H),$ but they have a very nonlocal relation with respect to the localized
operators. The wedge region is a borderline localization in that there exist
wedge-localized operators which \textit{even in the presence of interactions
create polarization-free one-particle} (and also multiparticle incoming)
\textit{states}. These operators are not identical to the incoming
creation/annihilation operators but they share the same Reeh-Schlieder domain
for wedge localization. The wedge restricted vacuum state is a KMS state and
the KMS property is intimately related to the one-step crossing \cite{foun}. A
proof of crossing from these localization properties would go beyond the more
modest aims of this paper.

Whereas in QM, relativistic or not, one has great liberty in manipulating
interaction potentials without leaving the setting of QM so that almost any
prescibed outcome can be accommodated, this is not the case in QFT. There the
locality principle is very restrictive and this tightness even show up in
theorems about the S-matrix as the Aks theorem saying that in a 4-dimensional
QFT nontrivial elastic scattering is not possible without the presence of
inelastic components \cite{BBS}. For the formfactors the statement
(\ref{form}) has a more popular stronger form which, although probably
provable, has according to my best knowledge presently the status of a fact
being supported by experience. This is the apparent validity of a kind of
benevolent Murphy's \textit{law: all couplings of local operators to other
channels (in the case of formfactors multiparticle channels) which are not
forbidden by superselection rules actually do occur.} Of course one needs to
"bang onto the vacuum", there is no "boiling soup" in a an inertial frame
without heating the "vacuum stove".

The formfactor aspect of a local operator is perhaps the best QFT illustration
of Murphy's law to particle physics. This tight coupling of channels through
the realization of the locality principle is both a blessing and a curse. It
attributes a holistic structure to QFT which on the one hand aggravates the
strategy to divide the difficult problem of (nonperturbative) model
construction into easier pieces, but on the other hand is the main reason why
this theory is much more fundamental than QM. In particular it does \ not
support the presently fashionable idea of "effective" QFT in which the
holistic aspect is largely ignored and which is eulogized when it does give a
wanted result and disregarded if it does not. Some interesting and pertinent
remarks about the importance of the holistic point of view in connection with
the problem of the energy density of the cosmological reference state can be
found in \cite{Ho-Wa}

Vacuum polarization as a concomitant phenomenon of QFT was discovered a long
time before the role of locality took the center stage. It is interesting to
reformulate Heisenberg's first observation in a more modern context by
defining partial charges by limiting the charged region with the help of
smooth test function. In Heisenberg's more formal setting the partial charge
of a free conserved current in a spatial volume V is defined as
\begin{align}
Q_{V}  &  =\int_{V}j_{0}(x,t)d^{3}x\label{Hei}\\
j_{\mu}(x,t)  &  =:\phi^{\ast}(x,t)\overset{\leftrightarrow}{\partial}_{\mu
}\phi(x,t):\nonumber
\end{align}
Introducing a momentum space cutoff, the norm of $Q_{V}\left\vert
0\right\rangle $ turns out to diverge quadratically which together with the
dimensionlessness of Q is tied to the area proportionality. Hence already on
the basis of a crude dimensional reasoning one finds an area proportionality
of vacuum polarization. The cutoff was the prize to pay for ignoring the
singular nature of the current which is really not an operator but rather an
operator-valued distribution.

The modern remedy is to take care of the divergence by treating the singular
current as an operator-valued distribution. Such calculations have been done
in the 60s \cite{Kast}\cite{Sti} by using spacetime test functions which
regularize the delta function at coalescing times and are equal to one inside
the ball with radius $R$ and fall off to zero smoothly between $R$ and
$R$+$\Delta R.$ Using the conservation law of the current one can then show
\cite{Kast} that the action of the regularized partial charge on the vacuum is
compressed to the shell ($R,R+\Delta R)$ and diverges quadratically with
$\Delta R\rightarrow0$ i.e. As expected, the vacuum fluctuations vanish weakly
as $R\rightarrow\infty$ (even strongly by enlarging the time smearing support
of $g$ together with $R$ \cite{Requ}) i.e. the limit converges independent of
the special test function weakly\footnote{Although the norm diverges, the
inner product of $Q_{R}\left\vert 0\right\rangle $ with localized states
converges to zero in compliance with the zero charge of the vacuum.} to the
global charge operator%
\begin{equation}
\lim_{R\rightarrow\infty}\int f_{R,\Delta R}(\vec{x})g(t)j_{0}(x,t)d^{4}x=Q
\end{equation}
With other words in the limit of global charges the vacuum polarization drops
out together with the test function dependence.

The interesting question in the context of the present section is the question
of what is the $R,\Delta R$ dependence when $\Delta R\rightarrow0.$ The answer
depends on the dimension of spacetime and the leading divergence can be
calculated for free currents in massless theories. The simplest case is that
of a chiral current which is localized on a light ray%
\begin{align}
&  \left\langle j(x)j(x^{\prime})\right\rangle \simeq\frac{1}{\left(
x-x^{\prime}+i\varepsilon\right)  ^{2}}\\
&  \left\Vert Q(g_{R,\Delta R})\Omega\right\Vert \symbol{126}\ln\frac
{R}{\Delta R}\nonumber
\end{align}
i.e. different from QM the dimensionless \textit{partial} charges diverges in
QFT, the first manifestation of vacuum polarization as first observed by
Heisenberg. In higher dimensional QFT the logarithmic behavior is modified by
powers in $\frac{R}{\Delta R};$ in particular in d=1+3 one obtains an
(logarithmically corrected) area law%
\begin{equation}
\left\Vert Q_{R,\Delta R}\Omega\right\Vert \symbol{126}(\frac{R}{\Delta
R})^{2}\ln\frac{R}{\Delta R} \label{Q}%
\end{equation}

The vacuum polarization-caused $\Delta R\rightarrow0$ divergencies of this
partial charge operator are preempted in a certain sense by the behavior of
the dimensionless localization-entropy; however despite similarities the
computation of the latter is conceptually more involved. The reason is that
the entropy is inherently nonlocal in the sense that it cannot be obtained by
a integrating a pointlike conserved current (or any other operator) but rather
encodes a holistic aspect of an entire algebra. Nevertheless the\textit{
splitting property} (for a description of its history see \cite{Haag}) is in a
certain sense the algebraic analog of the test-function smearing on individual
field operators.

Entropy in QM is an information theoretical concept which measures the degree
of entanglement. The standard situation is bipartite spatial subdivision of a
global system so that global pure states become entangled with respect to the
subdivision i.e. they can be written as a superposition of tensor product
states. The entropy is than a number computed according to the von Neumann
definition from the reduced impure state which results in the standard way
from averaging over the opposite component in one of the tensor factors.

The traditional quantum mechanical way to compute entanglement entropy was
applied to QFT of a halfspace (a Rindler wedge in spacetime) for a system of
free fields in a influential 1984 paper \cite{BKLS}. The authors started from
the assumption that the total Hilbert space factorizes in that belonging to
the halfspace QFT and its opposite. The calculation is ultraviolet divergent
and after introducing a momentum space cutoff $\kappa,$ the authors showed
that the cutoff dependence is consistent with an area behavior.%

\begin{equation}
S/A=C\kappa^{2} \label{cut}%
\end{equation}
where in the conformal case $C$ is a constant, $\kappa$ is a momentum space
cutoff and $S/A$ denotes the surface density of entropy. The method of
computation is again the integration over the degrees of freedom of the
complement region and the extraction of the entropy from the resulting reduced
density matrix state whose degree of impurity encodes the measure of the
inside/outside entanglement.

This calculation should be seen in analogy with Heisenberg's momentum space
cut-off calculation of vacuum polarization in the partial charge (\ref{Hei}).
In both cases the starting formula is morally correct but factually wrong.
Neither is the partial charge inside a region defined by a volume integral nor
do (as we know from discussions in previous sections) global states in QFT
permit an inside/outside factorization. These incorrect assumptions create the
divergencies which are then kept under the lid by the popular emergency kit of
QFT: momentum space cutoff. In both cases dimensional arguments lead to an
area proportionality (with logarithmic corrections possibly escaping the consideration).

The main advantage of the present spacetime approach versus a momentum space
cutoff argument is that the split property teaches us that the vacuum
polarization cloud hovers near the horizon in the split region characterized
by the sheet size $\Delta R$. The divergence for $\Delta R\rightarrow0$
indicates in no way a conceptual inconsistency or shortcoming of QFT which
must must be overcome with the help of quantum gravity. With other words the
localization entropy is a notion within a specified QFT, it does not need any
reference with respect to an ill-defined nonlocal "cutoff theory"\footnote{The
concept of a theory with a cutoff cannot even be defined in the presence of
interactions, even if one limits the construction to the family of soluble
factorizing models.}. There are simply some quantities whose sharp
localization causes divergencies but whose global value is perfectly finite;
for the global charge it is the finite value carried by one or several
particles and in the case of the entropy its global value in the ground state vanishes.

\section{The elusive concept of localization entropy}

Let us first apply the previously presented split idea to a two-dimensional
conformal QFT in which case the double cone is a two-dimensional spacetime
region consisting of the forward and backward causal shadow of a line of
length $L$ at $t=0$ sitting inside larger cone obtained by augmenting the
baseline on both sides by $\Delta L.$ As a result of the assumed conformal
invariance of the theory, the canonical split algebra inherits this invariance
and hence the entropy $Ent$ of the canonical split algebra can only be a
function of the cross ratio of the 4 points characterizing the split inclusion%
\begin{align}
Ent  &  =-tr\rho ln\rho=f(\frac{\left(  d-a\right)  \left(  c-b\right)
}{\left(  b-a\right)  \left(  d-c\right)  })\\
with~a  &  <b<c<d=-L-\Delta L<-L<L<L+\Delta L\nonumber
\end{align}
where for conceptual clarity we wrote the formula for generic position of 4
points. Our main interest is to determine the leading behavior of $f$ in the
limit $\Delta L\rightarrow0$ (two pairs of points coalesce) which is the
analog of the thermodynamic limit $V\rightarrow\infty$ for heat bath thermal systems.

The asymptotic estimate for $\Delta L\rightarrow0$ can be carried out with an
algebraic version of the \textit{replica trick} which uses the cyclic orbifold
construction in \cite{Lo-Fe}. First we write the entropy in the form
\begin{equation}
Ent=-\frac{d}{dn}tr\rho^{n}|_{n=1},~\rho\in M_{can}\subset\mathcal{A}(L+\Delta
L)
\end{equation}
Then one uses again the split property, this time to map the n-fold tensor
product of $\mathcal{A}(L+\Delta L)$ from the replica trick into the algebra
of the compactified line $\dot{R}=S^{1}$ with the help of the $n^{th}$ root
function $\sqrt[n]{z}.$ The part which is invariant under the cyclic
permutation of the n tensor factors defines the algebraic version \cite{Lo-Fe}
of the replica trick. The transformation properties under Moebius group are
now given in terms of the following subgroup of DiffS$^{1}$ written formally
as%
\begin{align}
&  \sqrt[n]{\frac{\alpha z^{n}+\beta}{\bar{\beta}z^{n}+\bar{\alpha}}},~L_{\pm
n}^{\prime}=\frac{1}{n}L_{\pm n},~L_{0}^{\prime}=L_{0}+\frac{n^{2}-1}{24n}c\\
&  \dim_{\min}=\frac{n^{2}-1}{24n}c\nonumber
\end{align}
where the first line is the natural embedding of the n-fold covering of Moeb
in diff(S$^{1})$ and the corresponding formula for the generators in terms of
the Virasoro generators. As a consequence the minimal $L_{0}^{\prime}$\ value
(spin, anomalous dimension) is the one in the second line. With this
additional information coming from representation theory we are able to
determine at least the singular behavior of $f$ for coalescing points
$b\rightarrow a,$ $d\rightarrow c$%
\begin{equation}
Ent_{sing}=-lim_{n\rightarrow1}\frac{d}{dn}\left[  \frac{(d-a)(c-b)}%
{(b-a)(d-c)}\right]  ^{\frac{n^{2}-1}{24n}}=\frac{c}{12}ln\frac{(d-a)(c-b)}%
{(b-a)(d-c)}%
\end{equation}
Since the function is only defined at integer n, one needs to invoke Carlson's
theorem. Apart from the split setting the calculation follows the same steps
as entropy calculations in condensed matter physics \cite{Cardy} which is
based on certain assumed properties of functional integrals which permit the
avoidance of momentum space cutoffs.

The resulting leading contribution to the entropy reads%
\begin{equation}
Ent_{sing}=\frac{c}{12}\ln\frac{(d-a)(c-b)}{(b-a)(d-c)}=\frac{c}{12}%
ln\frac{L(L+\Delta L)}{\left(  \Lambda L\right)  ^{2}}%
\end{equation}
where $c$ in typical cases is the Virasoro constant (which appears also in the
chiral holographic lightray projection).

This result was previously \cite{S2} obtained by the "inverse Unruh effect"
for chiral theories which is a theorem stating that for a conformal QFT on a
light-like line the KMS state obtained by restricting the vacuum to the
algebra of an interval is unitarily equivalent to a global heat bath
temperature state for a certain (geometry-dependent) value of the temperature.
The chiral inverse Unruh effect involves a change of length parametrization;
the length proportionality of the heat bath entropy (the well known volume
factor) is transformed into a logarithmic length measure. The inverse Unruh
effect has only been established in chiral QFT, but it points towards a
question of significant conceptual and philosophical importance: is there a
structural relation between heat bath and localization-caused thermal behavior
or are do they represent two unrelated physical phenomena ?

On expects the two monads to behave in the same way after reparametrizing in a
way which accounts for the different spacetime aspects of the two monads and
their different approximations by type I factors. In the thermodynamic case
the monad is approached by type I algebras of Gibbs states on systems in a box
of volume $V$ in the limit $V\rightarrow\infty~$whereas approximations of the
$\mathcal{A(O})$ monade is done by the type I factors obtained from the split
property. since the vacuum restricted to split type I factors also turn out to
be thermal (with respect to the modular Hamiltonian) one expects a
universality in the two kinds of thermal behavior. Therefore the relevant
question is: can the volume divergencies of the heat bath thermodynamic
entropy be set in relation to the $\Delta R\rightarrow0$ divergence in the
area behavior (possible modified by a logarithm) caused by vacuum polarization
as in (\ref{Q}) ? And do all dimensionless localized objects have the same
leading divergence for $\Delta R\rightarrow0.$ Since QFT does not know any
frame-independent position operator ("effective" substitutes have no place in
conceptual arguments) the question arises whether QFT can offer an analog to
the Heisenberg uncertainty relation. A universal relation between the leading
entropy/energy increase with the sharpness of localization is as close as one
could come. 

This thermal universality hypothesis would suggest the following
correspondence between the heat bath and the localization entropy%

\begin{align}
((kT)^{n-1}V_{n-1})_{T=2\pi}\simeq\frac{R^{n-2}}{\left(  \Delta R\right)
^{n-2}}ln\frac{R^{2}}{\left(  \Delta R\right)  ^{2}}  & \label{l}\\
Ent(h.b.)_{T=2\pi}=Ent(loc)  &
\end{align}
where the first line expresses the reparametrization of the dimensionless
(n-1)-volume factor in terms of a dimensionless logarithmically corrected
dimensionless area factor. Since localization thermality is a phenomenon of
modular theory which does not know anything about kT, the dimensionless area
is obtained from the thickness of the light slice which appeared already as a
$\Delta L$ in the logarithmic divergence for n=2, the case in which we
presented an rigorous proof based on the chiral inverse Unruh effect. The
above equality for the entropies means in particlar that the two matter
dependent finite constants in front of the leading divergencies are the same
if we use the same quantum matter for the heat bath and the localization
situation. For n%
$>$%
2 the relation is conjectural but its violation would cause serious problems
in our understanding of QFT. Naturally this correspondence can only be
expected for the leading term in the thermodynamic limit $V\rightarrow\infty$
respectively in the "funnel" limit $\Delta R\rightarrow0$ of decreasing split distance.

A mathematical proof would amount to the calculation of the von Neumann
entropy of the density matrix $\rho$ which results from the restriction of the
vacuum to the split tensor factor, a task which goes beyond the present
computational abilities in QFT. However it is possible to present some more
details supporting details about the geometrical aspects of the situation
which are closely related to the leading $\Delta R\rightarrow0$ behavior of
the dimensionless partial charge operators (\ref{Q}) caused by the vacuum
polarization cloud.

Compared with the chiral models in the beginning of this section which can be
controlled quite elegantly with the replica method, the question of higher
dimensional localization entropy looks more involved. In terms of inclusions
and relative commutants the funnel approximation to the double cone situation
is described in terms of the following split inclusion \cite{Do-Lo}
\begin{align}
&  \mathcal{A(D}(R))\subset\mathcal{N}\subset\mathcal{A(D}(R+\Delta R))\\
&  \mathcal{A}(ring)\equiv\mathcal{A(D}(R))^{\prime}\cap\mathcal{A(D}(R+\Delta
R)),~\nonumber\\
&  \mathcal{N=A(D}(R))\vee J_{ring}\mathcal{A(D}(R))J_{ring}\nonumber
\end{align}
where $\mathcal{N}$ is the canonically associated type I algebra in terms of
which there is tensor factorization as in (\ref{split}) and canonical means
that there is an explicit formula in terms of the double cone algebra
localized symmetrically around the orgin with radius $R$ and a larger one with
radius $R+\Delta R.$ The canonical formula for $\mathcal{N}$ is written in the
third line where $J_{ring}$ is the modular reflection for the ring algebra
defined in the second line. Note that this ring region is contained in a
\textit{light sheet} between the two horizons of $\mathcal{D}(R)$ and
$\mathcal{D}(R+\Delta R)$.

The crucial geometric input which leads to the desired result is the
realization that the relevant part for the area-like behavior is the fact that
the vacuum on $\mathcal{N}$ only contributes in the ring region since on
$\mathcal{D}(R)$ it is indistinguisable from the old vacuum. The ring region
is proportional to the area, and allowing for the previously established
logarithmic behavior in lightlike direction, one ends up with \cite{BMS}%

\begin{equation}
Ent(\mathcal{D}(R))\overset{\Delta R\rightarrow0}{\simeq}C(n)\frac{R^{n-2}%
}{\left(  \Delta R\right)  ^{n-2}}\frac{c}{12}ln\frac{R(R+\Delta R)}{\left(
\Delta R\right)  ^{2}},~C(0)=1
\end{equation}
where the logarithm is the only singularity in chiral conformal (n=2) models.
The naive geometrical argument would favor the dimensionless area law
involving the ring size $\Delta R$ without the logarithm, whereas the
\textit{presence of the logarithm can be viewed as representing a lightlike
length factor} which according to the chiral inverse Unruh effect is mapped
into a logarithmic divergence. In this way of counting there is a perfect
match with the (n-1)-volume factor apart form the fact that one length factor
has to be mapped into a logarithm.

The result contradicts the popular folklore that QFT is incomparible with an
area behavior, which is sometimes used delineate QFT in CST from QG. The
presence of the logarithm is important for our conjecture of thermal
universality \cite{BMS} which would find its most perfect expression in the
existence of a yet hypothetical higher dimensional inverse Unruh effect (more
in the concluding remarks); this remains an interesting problem for future research.

\section{Holography onto horizons, BMS symmetry enhancement}

The special role of null-surfaces as causal boundaries, which define places
around which vacuum polarization clouds form, suggests that there may be more
to expect if one only could make \textit{QFT on a light-front} a conceptually
and mathematically valid concept. That this can be indeed achieved is the
result of holography. Holography clarifies most of the problem which were
raised by its predecessor, the "lightcone quantization" and explains why the
older method failed. One of the reasons has to do with short distance behavior
since the naive restriction of fields to space- or light-like submanifolds
require the validity of the canonical quantization formalism i.e. a short
distance dimension not worse than sdd=1, even though the there is no such
restriction on the dimension of chiral fields living on a lightray.

However the causal localization principle in its algebraic formulation permits
to attach to each region the algebra of its causal shadow. For null-surfaces
the situation with respect to pointlike generators improves. In that case the
observable algebras indexed by regions on the lightfront are really pointlike
field-generated and the field generators are transversely extended chiral
observable fields $C(x,\mathbf{x})$ where $x$ denotes the lightlike coordinate
on the lightfront and $\mathbf{x}$ parametrizes the n-2 dimensional transverse
submanifold. The absence of transverse vacuum polarization would suggest to
expect their commutation relations to be of the form
\begin{equation}
\left[  C_{i}(x_{1},\mathbf{x}_{1}),C_{j}(x_{2},\mathbf{x}_{2})\right]
=\delta(\mathbf{x}_{1}-\mathbf{x}_{2})\sum_{k=0}^{m}\delta^{(k)}(x_{1}%
-x_{2})\check{C}_{k}(x_{1},\mathbf{x}_{1})\label{commut}%
\end{equation}
where the number m of operator contributions on the right depends on the scale
dimensions of the two operators on the left hand side. The transverse delta
function expresses the absence of transverse vacuum polarization which is a
rigorous-model independent result of the algebraic setting \cite{S1}. As for
standard chiral fields the scale dimensions are unlimited (no restriction to
canonicity as for equal time commutations)\footnote{There can be higher
derivatives in the transverse direction but they are always even whereas the
light-like delta functions are odd.}. The $\check{C}(x,\mathbf{x})$ denote
$\mathbf{x}$ dependent chiral fields of which has to know (as a consequence of
the absence of transverse vacuum fluctuations) only the product structure in
$x,~x^{\prime}$ at the same $\mathbf{x}$ which Such a commutation relation,
with the exception of d=1+1 where there is no transverse dependence, can
however not be quite correct for composite fields as a simple free field
calculation for $:A^{2}(x):$ shows \cite{Goe} in which case an ill-defined
square of a delta function appears (see below). This is perhaps a reminder
that one should not aim for the holographic projection of individual pointlike
fields in any literal sense, but rather seek pointlike generators of the
holographically projected algebra according to: bulk fields$\rightarrow$bulk
local algebras$\rightarrow$ holographic projection$\rightarrow$ construction
of pointlike generators.

The modular localization theory plays a crucial role in the construction of a
local net on the lightfront and its generating fields and for this reason one
must start with operator algebras which is in a standard position with respect
to the vacuum. Since the full lightfront algebra is identical to the global
algebra on Minkowski spacetime, one must start with a subregion on the
lightfront and the largest such region is half the lightfront whose causal
completion is the wedge (so that it can be seen as the wedge%
\'{}%
s causal (upper) horizon $H(W)$\footnote{This is the quantum version of causal
propagation with characteristic data on $H(W).$ A smaller region on LF does
not cast a causal shadow.})%
\begin{equation}
\mathcal{A}(W)=\mathcal{A}(H(W))\subset B(H)
\end{equation}
It is very important to avoid to project more into this equation than what is
actually written: this equality refers only to the position of the two
algebras within the full algebra $B(H);$ it does not refer to their local
substructure. The latter would be very different indeed; the local
substructure consisting of the net of (arbitrarily small) double cones inside
$\mathcal{A}(W)$ and that on $H(W)$ have no direct relation.

The local substructure on the horizon $\mathcal{A}(H(W))$ is obtained by
intersecting different $W$ algebras which have their horizons on the same
lightfront. In 4-dimensional Minkowski spacetime they are connected by a
7-parametric subgroup of the 10-parametric Poincar\'{e} group containing: 5
transformations which leave W invariant (the boost, 1 lightlike translation, 2
transverse translations, 1 transverse rotation) and 2 transformations which
change the edge of W (the two "translations" in Wigner's \textit{Little~Group}%
). This 7-parametric subgroup is precisely the invariance group of the
lightfront, but as a consequence of the absence of transverse vacuum
polarization of QFTs on null surfaces, the loss of symmetry is more than
compensated for by a gigantic symmetry gain leading to an infinite parametric
symmetry containing the Bondi-Metzner-Sachs group.

Although the net structure of the bulk determines that on a lightfront, the
inverse is not true, it is not possible to construct the net structure of
$\mathcal{A}(W)$ from that of $\mathcal{A}(H(W)).$ The additional information
beyond the intrinsic data of the $\mathcal{A}(H(W))$ net which will secure
unique inversion can have different appearance: Poincar\'{e} transformations
or characteristic propagation laws off $H(W)$ or the relative positioning
(forming a kind of \textit{algebraic GPS system}) of not more than three
lightfronts in different appropriate relative positions. The loss of
information, of phase space degrees of freedom and of symmetries (those which
transform out of the null surface) are all interconnected and related to the
projective nature of holography onto horizons a projection. The only known
case of a bona fide correspondence is the AdS$_{n}$-CFT$_{n-1}$ isomorphism in
which case the symmetry groups are identical.

For free fields the construction can be done explicitly. Since it is quite
interesting and sheds some light on why the holography works where the old
lightcone quantization did not succeed, the remainder of this section will be
used to present the free field holography\footnote{I am indebted to Henning
Rehren who informed me that similar idea can be traced back to work by Kay and
Wald from the 90s \cite{K-W}. A presentation of free field holography from a
more functional analytic point of view can be found in \cite{Da2} and
references therein.
\par
{}}.

The crucial property, which permits a direct holographic projection, is the
mass shell representation of a free scalar field%
\begin{equation}
A(x)=\frac{1}{\left(  2\pi\right)  ^{\frac{3}{2}}}\int(e^{ipx}a^{\ast}%
(p)\frac{d^{3}p}{2p_{0}}+h.c.) \label{A}%
\end{equation}
With the help of this representation one can directly pass to the lightfront
by using lightfront adapted coordinates \ $x_{\pm}=x^{0}\pm x^{3}%
,~\mathbf{x},$ in which the lightfront limit $x_{-}=0$ can be taken without
causing a divergence in the p-integration. Using a p-parametrization in terms
of the wedge-related hyperbolic angle $\theta:p_{\pm}=p^{0}+p^{3}\simeq
e^{\mp\theta},~\mathbf{p}$ the $x_{-}=0$ restriction of $A(x)$%

\begin{align}
&  A_{LF}(x_{+},\mathbf{x})\simeq\int\left(  e^{i(p_{-}(\theta)x_{+}%
+i\mathbf{px}}a^{\ast}(\theta,\mathbf{p})d\theta d\mathbf{p}+h.c.\right)
\label{LF}\\
&  \left\langle \partial_{x_{+}}A_{LF}(x_{+},\mathbf{x})\partial_{x\prime_{+}%
}A_{LF}(x_{+}^{\prime},\mathbf{x}^{\prime})\right\rangle \simeq\frac
{1}{\left(  x_{+}-x_{+}^{\prime}+i\varepsilon\right)  ^{2}}\cdot
\delta(\mathbf{x}-\mathbf{x}^{\prime})\nonumber\\
&  \left[  \partial_{x_{+}}A_{LF}(x_{+},\mathbf{x}),\partial_{x\prime_{+}%
}A_{LF}(x_{+}^{\prime},\mathbf{x}^{\prime})\right]  \simeq\delta^{\prime
}(x_{+}-x_{+}^{\prime})\delta(\mathbf{x}-\mathbf{x}^{\prime})\nonumber
\end{align}
The justification for this formal manipulation uses the fact that the
equivalence class of test function which have the same restriction $\tilde
{f}|_{H_{m}}$ to the mass hyperboloid of mass $m$ is mapped to a unique test
function $f_{LF}$ on the lightfront \cite{Dries}\cite{S1}. One easily verifies
the identity $A(f)=A(\left\{  f\right\}  )=A_{LF}(f_{LF}).$ But note also that
this identity does not mean that the \thinspace$A_{LF}$ generator can be used
to construct the localization structure from that of a characteristic initial
value problem which, concerning localization issues, is very different from a
Cauchy initial value problem. Even classically the lightfront-bulk relation is
primarily one between symplectic subspaces of the global symplectic space of
all classical waves, rather than relations between individual solutions.

The restriction to transverse or lightlike compact data does not improve the
localization within the wedge, it only causes "fuzziness", i.e. lack of
reconvertibility of an algebraic automorphism to a geometric diffeomorphism.
So algebraic holography from a wedge in the bulk to its horizon is only
invertible if one knows the law of characteristic propagation from the horizon
into the bulk; in the interaction free case this means the knowledge of the
bulk mass which was lost in the holographic projection. This law has no
geometric presentation, i.e. the local substructure of a wedge algebra
$\mathcal{A}(W)$ cannot be geometrically encoded into $\mathcal{A}(H(W)$),
although the two global algebras are identical. This also applies to event
horizons in curved spacetime; is incompatible with the idea that the
\textit{full} information contained in the local bulk substructure of a region
can be locally encoded into its horizon. 

In discussing the horizon-bulk relation it is easy to overlook the fact that
the representation of the lightfront generators in terms of the Wigner
creation and annihilation operators $a(p),a^{\ast}(p)$ as in the first line of
(\ref{LF}) is not intrinsic, rather the intrinsic characterization of the
theory is contained in the structure of it correlation functions or its
commutation relation. The characteristic data in the interaction free case
have lost all reference to a mass and unless one adds this information there
will be no unique holographic inversion. The uniqueness situation is in no way
better in interacting theories. Even in the free field situation the mutual
fuzziness between compact localized regions on $H(W)$ and regions in the
$W$-bulk or the inverse situation remains. In the spirit of LQP intrinsicness,
the reconstruction of the local substructure $\mathcal{A(}W)$ requires the
knowledge of the action of the Poincare group or that of the relative
positions of several null-surfaces. It turns out that the enlargement of group
symmetry beyond the 7-parametric subgroup and the increase of degrees of
freedom through the relative positioning are alternative ways for
reconstructing the bulk from it holographic projection (more remarks below).

Not knowing anything about QG, it is difficult to refute or support the claim
that there are holographic "screens" in QG which store \textit{all} bulk
information; but this is definitely not the situation in QFT on causal
horizons or on event horizons as they occur in QFT in CST. There are fewer
degrees of freedom in a QFT on LF than in the bulk QFT. More knowledge as that
of the action of LF-changing Poincar\'{e} transformations increases the
cardinality of degrees of freedom. The usefulness of the holographic LF
projection is inexorably linked to the thinning out of degrees of freedom. The
best way to appreciate this happy circumstance is to look at correspondences
for which this fails to be true (see below).

Attempts at nonperturbative constructions of QFT inevitably amount to
subdividing the problem into simpler pieces which only address certain aspects
of the holistic QFT project. Holography on horizons is a radical spacetime
reordering of a given quantum matter substrate. The latter may be a Weyl
algebra (the more rigorous formulation for (\ref{A}) and (\ref{LF})), a CAR
algebra, or that described in terms of a local interaction. In all cases the
spacetime reorganization according to the LF ordering structure simplifies
certain field aspects at the expense of particle aspects which become masked
(hidden in the holographic inversion for which more information is required).

Behind this idea of "thinning" degrees of freedom and loosing information in
holographic projection there is the concept of a \textit{natural phase space
density} for a given spacetime dimension. Intuitively speaking the idea behind
this is to hold onto as many properties as possible from Lagrangian QFTs in
situations outside the Lagrangian quantization setting. As mentioned in
section 2 of the part I this notion, introduced by Haag and Swieca
\cite{Ha-Sw} in the 60s and later refined to the nuclearity requirement by
Buchholz and Wichmann \cite{Bu-Wi} in the 80s, demands that the phase space
density of degrees of freedom in QFT, which is compatible with modular
localization, is bigger than the \textit{finite} degrees of freedom per phase
space cell of QM; but the infinite degrees of freedom also should not go
beyond that of a \textit{nuclear} set, since otherwise the causal propagation,
the existence finite temperature statistical mechanics and the asymptotic
particle interpretation will be endangered.

Such a situation arises in the AdS$_{5}$-CFT$_{4}$ correspondence because if
one choses one side, say the one with the larger spacetime dimension, as being
of Lagrangian origin (i.e. with a natural phase space density), the other side
of a correpondence is uniquely determined and the only thing one can do is to
look whether its degrees of freedom are natural or not. The naive argument
would suggest that when one passes to a lower dimensional world one has too
many degrees of freedom i.e. naturality is lost. In the opposite direction one
expects that the 5-dimensional AdS theory obtained from a natural CFT$_{4}$
model is too "anemic" the AdS theory coming from a normal CFT turns out to be
"anemic". Both statements can be made precise and exemplified by explicit free
field calculations starting from either side \cite{Du-Re}.

Strangly enough, although noticed very preciesly by Rehren, who gave a
mathematical proof of this correspondence\footnote{Something which is ill in
the physical-conceptual setting, maybe perfect on the mathematical side.}
\cite{Re1}\cite{Re2}, this issue has not been addressed by the Maldacena
community \cite{Mal} who first formulated the conjecture about the
correspondence in the context of a conjecture concerning the relation of
gravity on AdS$_{5}$ with a conformal supersymmetric N=4 Yang-Mills theory in
4 dimensions both thought of as theories with standard physical degrees of
freedom. There is a vast community with more than 6000 publications who tried
to support Maldacena's conjecture, but scientific truth are not decided
according to the size of globalized communities. In fact such communities
follow completely different pattern\footnote{In fact for the first time in the
history of particle theory there is a deep schism between a majority who has
been raised in the shadow of a theory of everything and a scholarly minority
with profound knowledge of QFT who are in a ivory tower against their own
choice. Particle theory has entered a deep crisis.} than a critical discourse
between individuals or small groups of individuals collaborating on one
subject. The conjecture has soon its 20$^{th}$ anniversary with no tangible
result but an ever larger number of publications with increasingly outragious claims.

The fact is, and every particle physicist of a sufficient age will confirm
this, that several decades of community building around the idea of a "theory
of everything" and extremely bad leadership has created an expectation of
salvation in which the level of knowledge falls far back behind what is needed
for research at the frontiers of QFT. 

Although not so obvious, the degrees of freedom argument can also be applied
to brane physics; against naive intuition branes contains the same cardinality
of degrees of freedom as the bulk\cite{foun} and hence the same arguments
about spacetime dimenion compatible naturalness applies.

We have seen that the holography of bulk matter on $W$ to the horizon $H(W)$
is not a correspondence but a projection. So it is clear that the loss of
information or the reduction of degrees of freedom for the preservation of
naturalness is a priviledge of holography on null-surfaces. This explains why
holography onto horizons is extremely useful. For the case at hand, namely the
bulk- and lightfront- generators, this projective nature of holography asserts
itself via the fact one cannot reconstruct the bulk from the space of $H(W).$
But the holographic projection is nevertheless very useful because it contains
still a lot of informations about the bulk in a much simpler more accessible
fashion. It is this aspect of simplification at the expense of information
completeness which makes holographic projection that useful. Of course this
could also happen in the case of correspondences; even though a CFT$_{4}$
viewed from the unphysical AdS$_{5}$ description has lost its physical
interpretation, certain mathematical aspects may still simplify.

Historically the "lightcone quantization" which preceded lightfront holography
shares with the latter part of the motivation, namely the idea that by using
lightlike directions one can simplify certain aspects of an interacting QFT.
But as the terminology "quantization" reveals that this was mixed up with the
erroneous idea that in order to achieve simplification one needs a new
quantization instead of a radical spacetime reordering of a given abstract
algebraic operator substrate whose Hilbert space is always maintained. As
often such views about QFT results from an insufficient appreciation of the
autonomy of the causal locality principle by not separating it sufficiently
from the contingency of individual pointlike fields.

Formally mass shell representations also exist for interacting fields. In fact
they appeared shortly after the formulation of LSZ scattering theory and they
were introduced in a paper by Glaser, Lehmann and Zimmermann \cite{GLZ} and
became known under their short name of "GLZ representations". They express the
interacting Heisenberg field as a power series in incoming (outgoing) free
fields. In case there is only one type of particles one has:
\begin{align}
&  A(x)=%
{\displaystyle\sum}
\frac{1}{n!}%
{\displaystyle\idotsint\limits_{V_{m}}}
a(p_{1},...p_{n})e^{i\sum p_{k}x}:A_{in}(p_{1})...A_{in}(p_{n}):\frac
{d^{3}p_{1}}{2p_{10}}...\frac{d^{3}p_{1}}{2p_{10}}\label{GLZ}\\
&  A_{in}(p)=a_{in}^{\ast}(p)~on~V_{m}^{+}~and~a_{in}(p)~on~V_{m}%
^{-}\nonumber\\
&  a(p_{1},...p_{n})_{p_{i}\in V_{m}^{+}}=\left\langle \Omega\left\vert
A(0)\right\vert p_{1},...p_{n}\right\rangle
\end{align}
where the integration extends over the forward and backward mass shell
$V_{m}^{\pm}\subset V_{m}$ and the product is Wick ordered. The coefficient
functions for all momenta on the forward mass shell $V_{m}^{+}$ are the vacuum
polarization components of $A$ and the various formfactors (matrix elements
between in "ket" and out "bra" states). In the GLZ setting the coefficient
functions arise as the mass shell boundary values of Fourier-transformed
retarded functions.

The convergence status of these series is unknown\footnote{In contrast to the
perturbative expansion which is known to diverge even in the Borel sense, the
convergence status of GLZ had not been settled.}. This mass-shell
representation is inherently nonlocal. Nevertheless one may hope that it does
not only represent a local bulk field but that its light front restriction is
also local. Superficially there is no problem with placing the GLZ
representation on the lightfront. However the application to $:A^{2}(x):$ the
Wick-ordered composite of the free field shows that there is an obstruction
against a simple-minded pointlike formulation (\ref{commut}) since the Wick
decomposition of $:A^{2}(x):_{LF}:A^{2}(x^{\prime}):_{LF}$contains squares of
transverse delta functions which, as the result of having lost the energy
momentum positivity in the transverse components, are incurably divergent.
This \textit{transverse delta problem} is absent in the holographic projection
of two-dimensional massive theories. The decisive property is however not
whether generating fields on LF come from pointwise manipulations on bulk
fields, but rather whether a net on LF can be described of generating fields.
But since the concrete calculations in terms of individual fields is more
familiar one would like to hope that there is a solution to the transverse
delta problem.

The holography on horizons contains some not entirely understood problems of
spin and statistics. Only Bose fields with integer short distance dimensions,
as those associated with conserved currents (conserved charge currents, the
energy-momentum tensor), can have bosonic holographic projections whereas
(bosonic or fermionic) bulk field with anomalous short distance dimensions
pass to plektonic lightfront fields for which the anomalous dimension, the
anomalous spin and their braid group statistics are interconnected via the
chiral spin\&statistics theorem \cite{Mund}. This change of the statistics in
passing from bosons/fermions with anomalous dimensions to lightfront fields
with anyonic/plektonic statistics is formally taken care of by the GLZ formula.

These transmutation properties with respect to statistics are more
conveniently studied in the simpler context of absence of transverse dimension
i.e. in the holographic projection of two-dimensional QFTs onto the lightray.
In this case the aforementioned obstruction is absent. In particular for the
factorizing models presented in the section on algebraic aspects of modular
theory, there are on-shell representation of local fields in terms of certain
wedge generating creation/annihilation operators, the Zamolodchikov-Faddeev
algebra generators (see part I), \ which replace the incoming
creation/annihilation operators in (\ref{GLZ}) and lead to a coefficient
functions which are identical to the crossing symmetric formfactors.

The pointlike fields in the mass shell representation highlight some
interesting problems whose better understanding is important for autonomous
nonperturbative constructions of models in QFT i.e. constructions which do not
depend on Lagrangian quantization as those presented in part I. The more
rigorous algebraic method by its very nature (using relative commutants) only
leads to bosonic holographic projections. This means that the extended chiral
structure on the lightfront only contains integral values in its short
distance spectrum; i.e. the generating fields are of the kind of the chiral
components of two-dimensional conserved currents and energy-momentum tensors.
Hence only a small subalgebra of the bulk algebra\footnote{Apart from
conserved currents whose charges must be dimensionless, fields are not
protected against carrying non-integer short distance scale dimensions.}
associated with transverse extended currents, energy momentum tensor etc. will
have a bosonic holographic image; there would be no anomalous dimension field
in the algebraic holographic projection. Apart from conserved currents whose
charges must be dimensionless, fields are not protected against carrying
non-integer short distance scale dimensions; such fields would not pass the
algebraic method of holography.

Clearly some of these ideas, as important for the future development of QFT as
they may appear, are not yet mature in the sense of mathematical physics.
Therefore it is good to know that there exists an excellent theoretical
laboratory to test such ideas in a better controlled mathematical setting, the
two-dimensional factorizing models and their this time bona fide (no
transverse extension) chiral holographic projection. From a previous section
on modular theory in part I one knows that these models have rather simple
on-shell wedge generators $Z(x)$ which still maintain a lot of similarity with
free fields. In that case Zamolodchikov proposed a consistency argument which
led to interesting constructive conjectures about relations between
factorizing models and their critical universality classes represented in form
of their conformal short distance limits.

From a conceptual viewpoint the critical conformal limit leading to
universality classes is very different from the holographic projection. The
former is a different theory whose Hilbert space has to be reconstructed from
the massless correlation function, whereas the latter keeps the original
Hilbert space and only reprocesses the spacetime ordering of the original
quantum substrate. Assuming that one knows the chiral fields on the lightray
as a power series in term of the Zamolodchikov-Faddeev operators\footnote{From
the point of view of chiral models such a representation is of course somewhat
unusual.} \cite{Z}, one has a unique inversion, i.e. the holographic
projection becomes an isomorphism.

Calculations on two models \cite{Ba-Ka}, the Ising field and the Sinh-Gordon
field, have shown that the universality class method and the holographic
projection lead to identical results\footnote{The consistency of the
holographic lightray projection with the critical limit for factorizing models
was verified in s blackboard discussion with Michael Karowski.}. Whereas the
anomalous dimension of the Sinh-Gordon field can only be computed
approximately in terms of doing the integrals in the lowest terms in the mass
shell contributions, the series for the Ising order field can be summed
exactly and yields the expected number 1/16. This is highly suggestive for
reinterpreting the more speculative Zamolodchikov way of relating factorizing
models with chiral models in the conceptually clearer setting of holographic projections.

The gain in modular generated symmetry is perhaps the most intriguing aspect
of holography. In general the modular theory for causally complete spacetime
regions smaller than wedges leads to algebraic modular groups which cannot be
encoded into diffeomeophisms of the underlying spacetime manifold; the
generators of these groups are at best pseudo-differential operators. However
there are strong indications that their restriction to the horizon are always
geometric. So it may be useful to construct the bulk modular groups from those
of their holographic projection. The constructive knowledge about chiral
theories has very much progressed \cite{Kawa} and it would be nice to be able
to use that insight to construct massive bulk theories with chiral models
being the holographic input.\ 

Let us finally address the symmetry enhancement which leads to the infinite
Bondi-Metzner-Sachs symmetry group which these authors discovered in
asymptotically flat solutions of classical general relativity. In the case of
the free field it is not difficult to see \cite{BMS} that the absence of
transverse vacuum polarization leads to a slightly larger symmetry than the
transverse Euclidean group; the transverse delta functions permits a
compactification to the Riemann sphere on whose complex $\zeta,\bar{\zeta}$
coordinates $(\zeta=x+iy)$ the group SL(2,C) acts as a fractional
transformation, just as the covering of the Lorentz group. Restricting the
Diff(S$^{1}$) group to the symmetry group of the vacuum which is the finite
parametric Moebius group; imposing in addition the requirement of the
preservation of the point at infinity in the lightlike direction the group is
the ax+b translation dilation group. By itself this would be a two parametric
group, but the fact that the two parameters can be functions of $\zeta
,\acute{\zeta}$ makes jointly generated group an infinite parameter group
\begin{align}
x  &  \rightarrow F_{\Lambda}(\zeta,\bar{\zeta})(x+b())\zeta,\bar{\zeta}\\
(\zeta,\bar{\zeta})  &  \rightarrow U(\Lambda)(\zeta,\bar{\zeta}%
),~U(\Lambda)\in SL(2,C)\nonumber
\end{align}
The group composition law $F_{\Lambda^{\prime}}(\Lambda(z,\bar{z}))F_{\Lambda
}(z,\bar{z})=F_{\Lambda^{\prime}\Lambda}(z,\bar{z})$ requires the
multiplicative factor to be of the form%
\begin{equation}
F_{\Lambda}(\zeta,\bar{\zeta})=\frac{1+\left\vert \zeta\right\vert ^{2}%
}{\left\vert a\zeta+b\right\vert ^{2}+\left\vert c\zeta+d\right\vert ^{2}}%
\end{equation}
whereas the functions $b(\zeta,\bar{\zeta})$ are from a function space which
is the closure of $C^{\infty}(\zeta,\bar{\zeta})$ functions on the Riemann
sphere in some topology. The somewhat unexpected property is that the action
of SL(2,C) on the function space contains (in its linear part) the a copy of
the semidirect product action of the Lorentz group on the translations i.e.
the infinite dimensional BMS group contains the Poincare group. For more
informations especially on the position of the Poincare inside the BMS group
we refer to a comprehensive paper by Dappiaggi \cite{Da2}.

One expects this transformation on a classical Penrose double cone horizon at
infinity since on such a "screen" the Poincar\'{e} group acts naturally. But
its appearance already on compact quantum double cones is at first sight
somewhat astonishing although the split property yields a mathematical
explanation \cite{BMS}.

It is helpful to take notice that in addition to the thermal property of the
vacuum reduced to one tensor factor as explained in the previous section, the
split property permits also to "localize" global symmetries which constitutes
an analog of the classical Noether theorem \cite{Haag}. This pure algebraic
derivation does not require to define a conserved current with the help of the
Lagrangian quantization, one even does not have to know how to construct
without being forced to postulate the existence of singular current operators
as the quantum counterparts of the classical conserved Noether currents. This
intrinsic (i.e. not relying on a quantization parallelism) "localization" of
global symmetries based on the split property also applies to the localization
changing Poincar\'{e} symmetry if one restrict the group parameters to
sufficiently small values so that the localization of the transformed
operators stays inside the chosen localization region \cite{Haag}.

Using the notation of the double cone localization defined in the previous
section one obtains a representation of the full Poincar\'{e} group on the
tensor factor $\mathcal{N}$ which for sufficiently small parameters act on
operators $A\in\mathcal{A(D}(R))$ the same way as the global symmetry. In the
ring region or its light sheet prolongation $\mathcal{D}(R)\backslash
D((R+\Delta R)),$ which constitutes the fuzzy localized part of $\mathcal{N}$
which surrounds its sharply localized nucleus $\mathcal{A(D}(R)),$ the
Poincare group does not act geometrically in a way which can be encoded into a
geometric diffeomorphism; of course it never fails to be an algebraic
automorphism. Hence the split situation for a double cone creates an analog
situation to a Penrose screen except that the Poincar\'{e} subgroup of the BMS
group is an unphysical extension of the partial physical Poincar\'{e} group
for parameter values for which the boundary of the region of interest in
$\mathcal{A(D}(R))~$passes into the $\Delta R$ split ring-like or light-sheet
region with $\Delta R\rightarrow0$ in the holographic limit in which
\textit{light-sheet }$\rightarrow$\textit{ holographic screen}$.$ With other
words the artifact a Poincar\'{e} subgroup of the holographic BMS group is
explained in terms of the artifact of a localized Poincar\'{e} symmetry
resulting from the split construction.

\section{The local covariance principle}

Less than a decade ago the holistic structure QFT in CST was significantly
enriched by the formulation of the local covariance principle \cite{BFV}.
Preliminary studies in this direction began at the beginning of the 90s with
the realization that even in the case of a free quantum field the definition
of an energy-stress tensor with properties similar to those of the classical
expression which enters the right hand side of the Einstein Hilbert equation
is a very nontrivial matter as soon as curvature enters \cite{Wa1994}. One
problem is that even in Minkowski QFT, where a unique definition in terms of
the Wick-ordered expression of the classical form is available, the energy
density is not bounded below, since one can find state vectors on which the
energy density $T_{00}(x)$ takes on arbitrarily large negative values
\cite{EGJ}. Whereas this result does not create serious problems in standard
QFT, it causes problems with the quantum counterpart of certain stability
theorems which follow from positivity inequalities for the classical stress
energy tensor which enters on the right hand side of the Einstein Hilbert equation.

It started a flurry of investigations which led to state-independent lower
bounds of $T_{00}(f)$ for fixed test functions as well as inequalities on
subspaces of test functions \cite{Ford}. These inequalities which involve the
free stress-energy tensor were then generalized to curved space
time\footnote{For recent publication with many references see \cite{Few}.}. In
the presence of curvature the main problem is that the correct definition of
$T_{\mu\nu}(x)$ is not obvious since in a generic spacetime there is no vacuum
like state (which is distinguished by its high symmetry) to which the operator
ordering could refer; to play that point split game with an arbitrarily chosen
state will not produce a locally covariant energy stress tensor since states
(in contradistinction to operators) are inevitably global in that their
dependence on the spacetime metric is not limited to the infinitesimal
surrounding of a point (which would be required by a local covariance principle).

A strategy to obtain locally covariant local quantum field product for the
energy-momentum tensor which is not associated with a particular state was
given in 1994 by Wald \cite{Wa1994} in the setting of free fields. His
postulates gave rise to what is nowadays referred to as the \textit{local
covariance principle} which is a very nontrivial implementation of Einstein's
classical covariance principle of GR to quantum matter in curved spacetime
(after freeing the classical principle from the relics of its physically empty
coordinate invariance interpretation). The requirements introduced by Wald
determines the correct energy-momentum tensor up to local curvature terms
(whose degree depends on the spin of the free fields). The method of Wald is
somewhat surprising since it does not consists in taking the coincidence limit
after subtracting from the point split expression the expectation value in one
of the states of the theory. Rather one needs to subtract a "Hadamard
parametrix" \cite{Wald2} i.e. a function which depends on a pair of
coordinates and is defined in geometric terms; in the limit of coalescence it
depends only on the metric in a neighborhood of the point of coalescence. Only
then the global dependence on the metric carried by states can be eliminated
in favor of a local covariant dependence on $g_{\mu\nu}(x)$ and its
derivatives. As a result the so-constructed stress-energy tensor at the point
$x$ depends only on the metric in an infinitesimal neighborhood of $x$.

Already Wald's work contains the important message that in order to construct
the correct tensor it is not enough to look at one model of a QFT in a
particular curved spacetime background, but one is obliged to look
\textit{simultaneously at all different spacetime orderings of abstract
quantum matter} (in Wald's case the abstract Weyl algebra quantum matter) in
order to be able to correctly describe the algebraic structure of one
particular model. The implementation of the local covariance principle
requires a strict separation of the algebraic structure from states; settings
of QFT in which the two are mixed together as functional integral approaches
or other formulations in terms of expectation values are unsuitable. In fact
it is not an exaggeration to think that without the dichotomy between
spacetime indexed nets of operator algebras and states inherent in algebraic
QFT, the formulation of QFT in CST would not reached the present level of clarity.

In \cite{BFV} the formulation of the local covariance principle attained its
present form. There are two different but connected formulations, one working
with nets of causally closed nets of spacetime-indexed operator algebras and
the other one in terms of pointlike covariant fields. The difference to the
standard formulation of a global algebra with its causally closed subalgebras
is that algebras which are "living" on isometric causally closed parts of
spacetime and are in addition algebraic isomorph (are made from the same
abstract matter substrate) are considered on equal footing. The totality of
observation which can be made on isometric isomorphic subalgebras is identical
and independent of differences which may show up in their surrounding. This
goes a long way towards what is considered as the characterizing property of
QG: the background independence. Some researchers of QG want to go one step
beyond isomorphy and look for equality in the spirit of gauge invariance by
integrating over gauge fields but a proposal to implement this ides is still missing.

The local covariance principle can also be expressed in terms of pointlike
covariant (under local isometries) fields. In contradistinction to standard
spacetime symmetries in QFT (e.g. Poincar\'{e} symmetry) these symmetries do
not come with a state which is left globally invariant. They are like the
Diff(S$^{1}$) symmetry beyond the Moebius group of chiral conformal QFT on the
circle in which case there is also no state which is globally invariant under
diffeomorphism beyond the Moebius group.

Recently these renormalization ideas were applied to computations of
backreactions of a scalar massive free quantum field in a spatially flat
Robertson-Walker model \cite{Bu-Mu-Su}. As a substitute for a vacuum state one
uses a state of the Hadamard form since these states fulfill a the so-called
microlocal spectrum condition which emulates the spectrum condition in
Minkowski spacetime. The singular part of a Hadamard state is determined by
the geometry of spacetime. The renormalization requirements of Wald lead to a
an energy momentum tensor with 2 free parameters which can be conveniently
represented as functional derivatives with respect to the metric of the two
quadratic invariants which one can form from the Ricci tensor and its trace.
In \cite{DFP} the resulting background equations were analyzed in the simpler
conformal limit and it was found that the quantum backreaction stabilizes
solutions i.e. accomplishes a task which usually is ascribed to the
phenomenological cosmological constant. Without the simplifying assumption the
linear dependence on a free renormalization parameter guaranties that any
measured value can be fitted to this backreaction computation. The principles
of QFT cannot determine renormalization parameters.

Hence from a QFT point of view there is no cosmological problem which places
QFT in contradiction with astrophysical observations. A consistency check
would only be possible if there are other measurable astrophysical quantities
which fall into the setting of quantum backreaction on spatially flat RW cosmologies.

Last not least the requirement of the local covariance principle to consider a
given quantum matter substrate simultaneously in \textit{all} CST helps to
maintain some aspects of particles, whose Wigner characterization only applies
to Minkowski spacetime. Since the latter is included in the covariance
definition it is sufficient to find an region of the given CST which is
isometric to a Minkowski space region in order to secure objects which behave
in a certain limited spacetime region as particles (see last section).

\section{Resum\'{e}, miscellaneous comments and outlook}

For a long time the conceptual differences between relativistic quantum
mechanics\footnote{In relativistic quantum mechanics (DPI of part I) the
velocity of light is not a maximal propagation over finite distances but
rather a limiting velocity for the leading asymptotic contribution of a wave
funtion. In this respect it is the relativistic counterpart of the speed of
sound in a nonrelativistic system of coupled oscillators.
\par
{}} and QFT in which the maximality of propagation is build into the algebraic
causality structure were not sufficiently appreciated. Even in contemporary
articles one finds the terminology "relativistic QM" instead of QFT. Perhaps
one reason is that many people believe that relativistic QM, as a separate
subject from QFT, does not really exist so that the somewhat sloppy
terminology does not really matter. But the existence of the DPI presented in
part I shows that this is not correct; the direct particle interaction theory
is a relativistic theory of particles which fulfills all requirement which one
is able to implement using exclusively properties of particles. As mentioned
in part I, even creation/annihilation processes of particles in scattering
processes can be described in DPI by introducing suitable channel couplings
"by hand". What is however characteristic of interacting QFT and has no place
in DPI is the notion of interaction-caused infinite vacuum polarization. In
part I the fundamental differences were explained in terms of two
fundamentally different localization concepts.

Fortunately these very different localizations coalesce asymptotically i.e.
the quantum mechanical Born-Newton-Wigner localization becomes covariant in
the asymptotic limit of scattering theory and its quantum mechanical
probability concept permits to extract cross sections from scattering
amplitudes. So perhaps it is better to de-emphasize the "bottle half-empty"
view expressed in (see part I) the title \textit{Reeh-Schlieder defeats
Born-Newton-Wigner} and take a more harmonic perspective by viewing
\textit{R-S and BNW, as an asymptotically harmonious pair.} Any other result
would have caused a disaster in the relation between particles and fields. DPI
reaches its conceptual limit if it comes to the notion of formfactors.

It is an interesting question whether LQP has any new message for the main
philosophical problem of the 20th century posed by QT: the controversy between
Bohr's (and more generally the Copenhagen) \textit{holistic view of quantum
reality} and Einstein's \textit{insistence in independent elements of
reality.} I think it does. On the one hand it pushes the holistic point of
view to its extreme as exemplified in the various ways it realizes an extreme
form of connectedness which we tried to highlight by calling it "Murphy's
theorem" ("what can couple does couple") as illustrated by Reeh-Schlieder
property, the analytic crossing connection of the different formfactors of a
local operator with its vacuum polarization and in a much stronger form by the
characterization of a LQP in terms of a finite number of monads in a specific
modular position. But on the other hand there is also the split property which
creates a situation close to Einstein's view. If one interprets Einstein's
maxim in an appropriate way, namely as the preservation of the totality of all
possible measurements in the presence of uncontrolled activities in a
spacelike separated laboratory instead of excluding the holistic EPR
situation, then there is no antagonism between the two views. The
reconciliation maybe difficult from an intuitive viewpoint, but the existence
of a mathematical consistent presentation clearly shows that intuition is not
always reliable and sometimes needs mathematical guidance.

The particle-field relation in the presence of interactions is one of the
subtlest aspect of relativistic local quantum physics; there has never been
any closure on this issue, nor would anybody who has a detailed knowledge
about this subject expect one in the near future. Nobody at a high energy
laboratory has ever directly measured a hadronic quantum field\footnote{There
are however certain distinguished composite fields, in particular the quantum
analogs of Noether currents, whose formfactors are used in the analysis of
scattering data for certain deep inelastic processes.}. Even though all our
intuition and the formulation of principles enters the theory through local
fields and spacetime indexed algebras of observables generated by them,
quantum fields remain hidden to direct observations. What one really measures
are either particles entering and leaving an interaction process, or thermal
radiation densities and their fluctuations as in the microwave background
radiation. Quantum fields or local observable algebras are the carriers of the
causal locality principle\footnote{In the noninteracting case covered by
Wigner's representation theory this viewpoint has led to the understanding of
string-localized generators of "infinite spin" representations \cite{MSY}.}
but, different from classical relativistic fields which propagate with a
maximum velocity, they have themselves no ontological status. The protagonists
of LSZ scattering theory coined a very appropriate word for this state of
affairs, they called fields in particle physics "interpolating". In general
there will be infinitely many interpolating fields which interpolate the same
particle. But there are indications based on the use of the crossing property,
that the inverse scattering problem has a unique solution with respect to the
system of local algebras \cite{inverse} without any guaranty for its existence.

Besides the standard Wigner particle setting whose connection with fields is
channeled through the (LSZ, Haag-Ruelle) scattering theory there are charged
(infra)particles whose scattering theory in terms of inclusive cross sections
exists in the form of computational recipes without conceptual backup. These
particle-like objects correspond to charged fields in QED which only exist as
semiinfinite strings i.e. are nonlocal (in the standard use of this word where
everything which is not pointlike generated is called nonlocal. It would be
naive to expect that the situation with respect to the necessity of
introducing physical nonlocal observables decreases in passing from abelian
gauge theories to Yang-Mills theories.

But it is precisely the idea of an equivalence class of interpolating local
fields in their property of interpolating the same particle which led to the
powerful observed properties as e.g. Kramers-Kronig dispersion relations which
a particle-based approach as DPI can not deliver. The experimental
verification of a dispersion relation cannot select or rule out a particular
Lagrangian model of hadronic interactions but rather is a test for the
validity of the causal localization principle.

The particle based view is certainly thrown into disarray when one studies QFT
in non inertial frames (e.g. the Rindler frame of the Unruh effect) or in CST.
According to the best of my knowledge there exists no time-dependent LSZ
scattering in a (flat) Rindler world; although the global and the
wedge-localized QFT live in the same Hilbert space, the global particle states
carry no intrinsic physical information with respect to the wedge-localized
theory. Robert Wald, a leading researcher on QFT in CST, has recently proposed
\cite{Wald2} to \ abandon the particle concept altogether and work under the
hypothesis that fields are directly measurable. But measurability requires a
certain amount of stability and individuality; quantum fields are fleeting
objects of which there are always infinitely many for which, in
contradistinction to classical fields, there seems to exist no measurable
property which allows to distinguish the members in an equivalence class of
fields which carry the same charge. It is hard to believe how Wald's advise of
abandoning particles could work. Perhaps, as indicated at the end of the
previous section, the local covariance law leads to an argument why certain
particle manifestations in Minkowski spacetime can be transfered to finite
regions in CST.

In recognition of this lack of observational distinctness for fields, the
algebraic approach to QFT has placed \textit{spacetime-indexed operator
algebras} into the center stage. In such a setting the increase of knowledge
about a localized operator algebra takes place through a tightening in
localization and not via the increase in precision in measuring properties of
an individual operator\footnote{An exception are those localized individual
operators which result from the "split localization" of global symmetries (the
before mentioned quantum Noether currents).}. This fits very nicely with
scattering theory because the in/out fields resulting from different operators
in the same algebra $\mathcal{A(O})$ are identical; their differences become
absorbed into normalization factors \cite{Araki} and it is at best the system
of operator algebras which is determined by inverse scattering and never the
individual fields.

Although without the notion of particles and scattering theory the physical
world of QFT in CST would be quite a bit poorer, it is by no means void of all
experimentally accessible aspects. Even if one has no clear idea on the nature
of the cosmological reference state of our universe (the CST replacement for
the vacuum), one can study models and compare the thermal aspects of the
expectation values of the energy-stress tensor in the cosmic reference state
with data from the cosmic background radiation \cite{DFP}, for this one does
not need the vacuum state and particle states as they follow from Poincar\'{e} symmetry.

A class of objects between particles (on-shell) and fields (off-shell) which
are ideally suited for the study of vacuum polarization are the
\textit{formfactors}, i.e. matrix-elements of local operators between
\textit{bra} out- and \textit{ket} in- particle states. The special
matrix-elements with vacuum on one side and all particles on the other side
characterize the vacuum polarization of the local "bang" on the vacuum
$A\Omega,~A\in\mathcal{A(O}).$ The general formfactor results from the vacuum
polarization component by a particular on-shell analytic continuation process
known as the \textit{crossing property.} The latter is one of the most subtle
property in the particle-field relation \cite{foun}, its comprehension goes
significantly beyond that of time-dependent scattering theory. As the Unruh
effect, it uses KMS properties of the wedge localized algebra\footnote{It
testifies to the conceptual depth of modular localization that it places such
diverse looking issues as the Unruh effect and the crossing property under one
roof.}, the subtle point being that one needs to construct very special wedge
localized operators which applied to the vacuum generate particle states
without admixture of vacuum polarization cloud \cite{foun}. The history of the
crossing property is also a prime example of the disastrous consequences of a
several decades lasting misunderstanding of a central concept of QFT
\cite{foun} in the absence of a profound criticism.

An example of a deep antagonism of QFT with respect to QM which is usually not
perceived as such comes from the exploration of modular localization. Whereas
the localization of states is basically a kinematical notion, its algebraic
version incorporates most, if not all dynamics. The crucial property is the
monad (hyperfinite type III$_{1}$ factor) nature of the local algebras. In QM
all Born-localized subalgebras are of the same type as the global algebra,
namely type I$_{\infty}$ factor $B(H),$ $H\subset H_{glob}.$ A monad in QM
only appears at finite temperature in the thermodynamic limit. There is hardly
any textbook which emphasises the radically different algebraic properties
(see however \cite{Robin}) from those of its "boxed" Gibbs state
approximands\footnote{The thermofield formalism of doubling of degrees of
freedom holds for the finite box and corresponds to the tensor factorization
between the boxed algebra and its commutant. But by not noting that this
factorization breaks down in the thermodynamic limit the aficionados of
thermofield theory miss an interesting chance of becoming aware of a deep
conceptual problem.}.

In QFT as opposed to QM, it is the monad structure which is the normal
situation and the quantum mechanical type I$_{\infty}$ property which is the
exception; the latter can only be constructed by "splitting" a local algebra
from its causal disjoint and in this way creating a fuzzy "halo" in which the
vacuum polarization can settle down to a (halo-dependent) temperate behavior
leading to a finite (halo-dependent) entropy. So the region for the
calculation of entropy is not the horizon itself but rather a light-sheet
surrounding the horizon of the localization region. Hence the divergence of
localization entropy in the limit of vanishing sheet size $\Delta
R\rightarrow0$ is not an indication that QG must intervene in order to rescue
QFT from high energy inconsistencies \cite{BKLS}, but rather that the
assumption of tensor factorization, which is the prerequisite of a bipartite
entanglement situation, was not quite correct; the total algebra
$B(H)=\mathcal{A}\vee\mathcal{A}^{\prime}$ cannot be written as a tensor
product even though $\mathcal{A}^{^{\prime}}$ is the commutant of
$\mathcal{A}$.

The split construction enforces the tensor product situation but it brings a
new parameter into the fray, the split size $\Delta R.$ The conceptual
situation calls for great care in using standard notions of quantum
information theory from QM in quantum field theoretical situations in which
thermal aspects of entanglement (and not the information theoretical) are
dominant. In particular the discussions about information loss in black hole
physics seems to have been carried out without much appreciation for the field
theoretical subtleties addressed in this essay. Although the terminology
"entanglement" strictly speaking does not apply to a bipartite separation with
sharp causal boundaries in QFT, the literature on entanglement unfortunately
does not seem to differentiate between the QM and the QFT case. There is of
course the problem of respecting a historically accepted terminology when its
literal meaning contradicts mathematical facts \cite{Keyl-M}.

Another issue presented in part I is the question to what extend one needs to
go beyond pointlike generators. We reviewed the Wigner representation theory
in the modular localization setting and reminded the reader that the only
class which needs stringlike generating covariant wave function is Wigner's
infinite spin class which after a more than 60 year odyssey, thanks to modular
localization, finally reached its final position with respect to localization.
The algebraic notion of stringlike generator of an algebra is however more
restrictive in the sense that it is described by an "indecompoaible"
string-like localized field $\Phi(x,e)$ (with $e$ the spacelike direction of
the semiinfinite string and x its start) i.e. one which cannot be resolved in
terms of a pointlike field smeared along $x+\mathbb{R}_{+}e.$ The application
of such an algebraic string to the vacuum state is however decomposable as a
state into irreducible representations of the Poincar\'{e} group and unless
there are infinite spin components, the localization of the state is pointlike
even though the algebraic object was an indecomposable string\footnote{The
state localization structure is exclusively determind by the representation
theory of the Poincar\'{e} group whereas the problem of irreducible algebraic
generators in interacting theories depends on the dynamics.}. The only
illustration for such an algebraic string, mentioned in the first part, is the
Dirac-Jordan-Mandelstam string. The Buchholz-Fredenhagen setting \cite{B-F}
offers room for pure massive strings as one wants them in QCD, but since these
strings do not leave any traces in perturbation theory, they remain beyond
what one is able to control with existing methods. We also stressed that the
widening of the setting of localization achieved through the modular formalism
poses new questions involving massive vectormesons whose resolution could be
relevant for the interpretation of forthcoming LHC experiments.

Localization is the overriding principle of LQP, in fact it is the only
principle and therefore the main and often difficult task in the conceptional
conquest of specific effects and mechanisms in QFT consists in figuring out
how and under what conditions they can be derived from localization. In
phenomena first observed in a Lagrangian quantization setting as e.g. QED
infrared properties, spontaneous symmetry-breaking a la Goldstone, the
Schwinger-Higgs screening mechanism or the crossing property most of the rich
conceptual-mathematical understanding came from the pursuit of this goal, and
if there has not yet been a perfect understanding, it only means the the
subtle connection of these phenomena with localization has not yet been
completely unraveled. The story of the Wigner infinite spin representation
class shows that even for kinematical problems the understanding of their
localization aspects sometimes take more than half a century (see part I). 

Even where it is least expected, namely in case of the mysterious quantum
concept of \textit{internal symmetries} is a particular mode of realization of
the locality principle. The DHR theory showed how the possible superselected
representations of an observable algebra\footnote{This example (the
Doplicher-Haag Roberts superselection theory \cite{Haag}) is particularily
suitable since nobody would expect group theory to emerge from classifying
inequivalent local representations of the observable algebra; at no point is
group theory visible.} and the ensuing group theory which ties the different
sectors into one "field representation" (on which it acts in such a way that
the local observable algebra reemerges as the fix point subalgebra under the
action of a compact group) is uniquely contained in the structure of the local
observables. On the other hand the monad picture shows that even spacetime
symmetries can be encoded in the abstract modular positioning within a shared
Hilbert space.

A brief explanation of this last remark is as follows (see part I). A QFT in
the algebraic setting is a net of spacetime-indexed algebras. Hence it comes
as somewhat of a surprise that one can do with less; to get a concrete QFT
going, one only needs a finite number of monads in a special relative modular
position. The reason why I used a whole section in part I of may essay for a
description of this property (which up to now has remained without practical
use), is that I find this very exciting from a philosophical point of view. It
is the almost literal adaptation of Leibniz's idea of what constitutes reality
to the setting of local quantum physics. A monad in isolation is not much more
than a point in geometry, besides the absence of pure and mixed states and the
statement about what kind of states it admits instead it, is an object without
properties. Surprisingly its is not even necessary to require that the
algebras are monads, their modular theory together with the positioning
defined in terms of modular inclusions or modular intersections \cite{Wi}
alone forces the factors to be of hyperfinite type III$_{1},~$no other factor
algebras can be brought into that particular modular position. The structural
richness of QFT result solely from the relation between these monads; this
includes not only the local net of quantum matter but also its internal as
well as spacetime symmetries. There are other equivalent ways to characterize
a QFT in terms of the modular data of its local subalgebras \cite{CGMA}. From
a practical point of view it turns out to be more useful to know the action of
the Poincare group on the generators on one fixed wedge which is equivalent to
the positioning point of view, indeed this was the approach by which the
existence of certain two-dimensional factorizing models was established
(section 5 of part I).

Perhaps the most profound conceptual-philosophical contrast between QM and QFT
finds its expression in these modular encoding. As mentioned in the last
section in part I, there have been other ideas which are designed to highlight
an underlying relational nature of QT; in particular Mermin's view of QM
\cite{Mermin} in terms of its \textit{correlations}. It is difficult to find a
mathematical backup which is as crystal clear as that in terms of modular
positioning of monads. Mermin expressed his relational point of view by the
following apodiction: \textit{correlations have physical reality, that what
they correlate does not.}

The LQP analog of this dictum would be: \textit{relative modular positions in
Hilbert space have physical reality, the substrate\footnote{Modular
positioning is the most radical form of relationalism since the local quantum
matter arises together with internal and spacetime symmetries. In other words
the concrete spacetime ordering is preempted in the abstract modular
positioning of the monads in the joint Hilbert space. } which is being
positioned does not.}

The presentation of QFT in terms of positioning monads is very specific of LOP
i.e. it has no analog in QM i.e. Mermin's relational view is not a special
case of positioning in LQP. It has the additional advantage that beyond the
metaphor there are hard mathematical facts.

Of course it would be a serious limitation if this philosophical viewpoint is
restricted to the characterization of Minkowski spacetime QFT. Despite all the
progress with QFT in CST in relation with the formulation of the local
covariance principle, it is too early for such questions involving modular
theory. As a preliminary test one could ask whether the diffeomorphisms in
Diff(S$^{1}$) beyond the Moebius transformations in a chiral theory, which as
well-known do not leave the vacuum state invariant, have their origin in
modular theory. It is clear that in order to achieve this one has to be more
flexible with states, i.e. using also other than the vacuum state and not
insisting in a modular automorphism of being globally geometric. In this
connection it is very encouraging that recently the idea of local covariance
found a satisfactory expression; without having a precise description of this
crucial principle, there would not be much chance to make headway with modular
localization methods in the CST setting of QFT.

The restriction of globally pure state (vacuum, particle states) to causally
localized subalgebras $\mathcal{A(O})$ leads to thermal KMS states associated
with the modular Hamiltonian associated to ($\mathcal{A(O}),\Omega$). Modular
Hamiltonians give rarely rise to geometric movements (diffeomorphisms).
Although outside conformal QFT there is no compact localization region in
Minkowski spacetime which leads to a fully \textit{geometric} modular theory,
there are rather convincing arguments that the modular automorphism becomes
geometric at the horizon of the localization region. The reason is that the
holographic projection onto the horizon is a (transverse extended) chiral theory.

In case of timelike Killing symmetries in CST there may even exist an
extensions of the spacetime and a state on it such that the modular group of
its restriction is identical to the Killing group\footnote{The standard
example is the Hartle-Hawking state on the Kruskal extension restricted to the
region outside a black hole.}.

One of the most mysterious aspects of localization-caused thermal behavior is
its possible connection to ordinary i.e. heat bath thermality. The problem is
often referred to as the inverse Unruh effect because one wants to know
whether there exists a heat bath thermal system which can be viewed as arising
through restricting the vacuum on an extended system i.e. in the spirit of an
Unruh effect. This is indeed possible for chiral theories and the connection
between the two systems is a conformal transformation which maps the standard
volume (here length) factor into the logarithm of the splitting length.
Although in higher dimension there is no rigorous argument which relates the
two kinds of thermal behavior, there is an educated guess.

Apart from the case of chiral theories on the lightray, where the validity of
the \textit{inverse Unruh effect} permits to transform the heat bath volume
(here length) law into a logarithmic divergence in $\Delta R\rightarrow0,$ a
relation between heat bath- and localization-caused entropy is unknown.$~$For
higher dimensional localization entropy the split property suggests to
consider the entropy of a light sheet whose thickness is the split distance
$\Delta R.$ By generalizing the localization entropy of chiral theories and by
relying on the vacuum polarization divergence of a dimensionless charge we
presented a formula for the leading divergence of the localization entropy and
gave strong arguments in favor of a universality between heat bath and
localization-caused entropy. From an algebraic viewpoint one parametrizes the
approximation of a monad by type I algebras in two different way; in the
thermodynamic limit by a $L^{3}$ proportional sequence and in the funnel limit
one of the length factors was replaces by a log so that effectively one
obtains a $L^{2}$ (area) proportionality. The monade itself is structureless
and the parametrization only appears in the context if its use inside a
physical theory.

A $\Delta R$ independent area law as that of Bekenstein, in which the light
sheet width $\Delta R$ is replaced by the Planck length, is not compatible
with the localization structure of QFT which requires the quadratic increase.
There is a tight connection between modular localization and the phasespace
density of states. Whereas the phase space density in QM is finite, that of
QFT is \textit{nuclear. }If one interprets the Bekenstein area law as coming
from a future quantum theory of gravity (QG) without standard quantum matter,
the algebraic structure of such a theory must be that of a very low phasespace
density, as e.g. an unknown QM or a combinatorial algebra with trace states.
On the other hand the localization entropy from QFT is the precise entropical
counterpart within the thermal setting of the Hawking radiation which does not
need any appeal to a yet unknown QG. Hence their remains a basic clash between
Hawking radiation, which Hawking derived from QFT localized outside a black
hole horizon, and the Bekenstein formula which was inferred from interpreting
a certain classical area formula.  

In the article we also presented examples of unnatural QFT in which the phase
space density is too high or too low. Among the unphysical consequences are:
the existence of a Hagedorn temperature or the absence of any thermal state,
as well as serious problems with causal propagation. In particular they cannot
occur in causally propagating situations as formally described by Lagrangian
quantization. The algebraic approach to QFT from its very beginning
\cite{H-S}\cite{Ha-Sw} tried to isolate them. Their only use has been to
exemplify those unphysical properties which a non-Lagrangian approach must
avoid and understand those properties which one must require to exclude
pathologies. They occur in infinite component QFT as string
theories\footnote{Contrary to its name, the result of the canonical
quantization of the Nambu-Goto Lagrangian is not string-localized but
represents a point-localized \textit{dynamical infinite component QFT}
\cite{foun}\cite{interI}.}. One also meets them in the AdS$_{n}$-CFT$_{n-1}%
~$correspondence, an explicit illustration is provided by taking a free
massive AdS field which on the conformal side yields a generalized conformal
field with the mentioned pathologies. Unfortunately the old insights into what
constitutes a natural QFT outside the Lagrangian protection have been lost on
protagonists of the supersymmetric N=4 Yang Mills -- super gravitation AdS
model. The attempt to remind them of the problems in their conjectures has
remained without avail \cite{Du-Re}.

The Bekenstein thermodynamical interpretation of a certain quantity in the
setting of classical gravity raises the question whether it is not possible to
invert this connection i.e. to supplement the thermodynamical setting by
reasonable assumptions of a general geometric nature, so that the Einstein
Hilbert equations are a consequence of the fundamental laws of thermodynamics.
Modular theory already relates thermal behavior with localization, hence a
relation of fundamental laws of thermodynamics with gravity is not as
unexpected as it looks at first sight. The reader is referred to some very
interesting observations by Jacobson \cite{Jac}.

Whereas there is hardly any doubt that apart from problems of improved
formulations the QM-QFT interface had reached its conceptual final position
this is not the case with the interface between QFT in CST and QG. Up to
recently the general belief was that the background independence and the
entropical area proportionality are marking this interface. But in both cases
this had to be amended. On the one hand the new local covariance principle
shows that local covariance implies at least the unitary quantum equivalence
of QFT in spacetime regions which are isometric\footnote{It is not clear
whether the stronger form of background independence, in which the isomorphism
is replaced by an identity, can be achieved.} which is a big step in the
direction of background independence. And if one ignores the logarithmic
factor the area proportionality by itself cannot be characteristic for QG. 

Returning to the main point in part I; there are hardly two concepts which are
that different than relativistic QM and relativistic QFT. In textbooks this is
consistently overlooked probably as a result of believing that because they
share the Lagrangian quantization formalism and $\hslash$ the only difference
is taken care of by adding the word "relativistic". In both parts of this work
we explained the difference in terms of the different localization concepts
which in turn is intimately related to the difference in the cardinality of
phase space degrees of freedom (finite in QM, "nuclear" in QFT \cite{foun}).
The ignoration or misunderstandings of these differences has been the cause of
a major derailment of particle physics \cite{foun}: string theory and the
Maldacena conjecture. But modular localization also led to the first existence
proof for certain interacting field theories (factorizing models) after 80
years of QFT, and it promises to revolutionize gauge theories \cite{char}. In
addition it generates the concepts which are necessary two "split" causally
separated regions so that the notion of entanglement can also be used in QFT
where it has thermal consequences (localization-caused thermal behavior).

Since the crisis of particle physics originated from confusing the
\textit{holistic} aspects of modular localization in QFT \cite{foun}%
\cite{interI}\cite{Jor} in the aftermath of S-matrix theory in the 60's and
has solidified ever since, creating a quite misleading intuition even in
present day QFT, the only way out is to correct this incorrect understanding
of the most important concept which constitutes the essence of QFT. This would
have been possible in earlier times when particle physics was done by
individuals or represented "schools" of thought. Whether the correction of
something which for several decades had the blessing of globalized communities
and has already solidified is possible et all, remains to be seen.

\end{document}